\documentclass[conference]{IEEEtran}
\IEEEoverridecommandlockouts
\usepackage{algorithm}
\usepackage{algorithmicx}
\usepackage{algpseudocode} 
\usepackage{enumitem}
\newenvironment{Proof}{{\it Proof:}\quad}{\hfill $\blacksquare$\par}

\usepackage{cite}
\usepackage{amsmath,amsthm,amsfonts,amssymb}
\usepackage{graphicx}
\usepackage{textcomp}
\usepackage{xcolor}
\def\BibTeX{{\rm B\kern-.05em{\sc i\kern-.025em b}\kern-.08em
    T\kern-.1667em\lower.7ex\hbox{E}\kern-.125emX}}

\theoremstyle{plain}
\newtheorem{thm}{Theorem}

\newtheorem{prop}[thm]{Proposition}

\allowdisplaybreaks[4]

\begin{document}

\title{Finite-Alphabet-Aware Trajectory and Precoder Optimization for UAV Relaying
}

\author{
Haoyang Di,
Xiaodong Zhu,
Yulin Shao
\thanks{H. Di and X. Zhu are with the School of Information and Communication Engineering, University of Electronic Science and Technology of China, Chengdu, China (e-mail: dhy@alu.uestc.edu.cn, zxdong@uestc.edu.cn). 
Y. Shao is with the State Key Laboratory of Internet of Things for Smart City and the Department of Electrical and Computer Engineering, University of Macau, Macau S.A.R. He is also with the Department of Electrical and Electronic Engineering, Imperial College London (e-mail: ylshao@ieee.org).
}
}

% \author{\IEEEauthorblockN{1\textsuperscript{st} Haoyang Di}
% 	\IEEEauthorblockA{\textit{SKL-IOTSC} \\
% 		\textit{University of Macau}\\
% 		Macau, China \\
% 		dhy@alu.uestc.edu.cn}
% 	\and
% 	\IEEEauthorblockN{2\textsuperscript{nd} Xiaodong Zhu}
% 	\IEEEauthorblockA{\textit{School of Information and Communication Engineering} \\
% 		\textit{University of Electronic Science and Technology of China}\\
% 		Chengdu, China \\
% 		zxdong@uestc.edu.cn}
% 	\and
% 	\IEEEauthorblockN{3\textsuperscript{rd} Yulin Shao}
% 	\IEEEauthorblockA{\textit{SKL-IOTSC} \\
% 		\textit{University of Macau}\\
% 		Macau, China \\
% 		ylshao@um.edu.mo}
% }

\maketitle

\begin{abstract}
Unmanned aerial vehicles (UAVs) have become key enablers in relay-assisted wireless communications thanks to their flexibility and line-of-sight channel advantage. However, most existing trajectory optimization frameworks assume ideal Gaussian inputs, overlooking the fact that practical wireless systems rely on structured, finite-alphabet constellations. This mismatch can lead to suboptimal, and sometimes misleading, design choices. In this paper, we challenge that convention by introducing a finite-alphabet-aware framework for joint trajectory and precoder optimization in UAV-assisted relay systems. We formulate a non-convex design problem that directly accounts for discrete signal structures and propose an efficient solution based on alternating optimization and successive convex approximation. Simulation results reveal that strategies optimized under Gaussian assumptions can waste energy and degrade throughput in real deployments. In contrast, our approach adapts both the UAV's trajectory and transmission strategy to the underlying modulation format, delivering consistent performance gains under practical system constraints. This work takes a key step toward aligning UAV communication design with the realities of modern wireless systems: discrete signals, power limits, and intelligent mobility.
\end{abstract}


\section{Introduction}
Unmanned aerial vehicles (UAVs) are rapidly transforming wireless communication, offering unparalleled agility and high mobility \cite{1,2}. Their ability to dynamically reposition, bypass obstacles, and establish reliable line-of-sight links makes them particularly valuable for relay-assisted communications, especially in challenging or emergency scenarios where terrestrial infrastructure is unavailable or compromised \cite{3,shao2021federated}.

To improve throughput and coverage in such UAV-assisted relay systems, extensive research has explored the joint optimization of beamforming and trajectory design \cite{4,5,6,7,8}. 
The authors of \cite{5}, for example, investigated a UAV-assisted relay system enhanced by an intelligent reflecting surface (IRS), aiming to jointly maximize spectral efficiency and energy efficiency. They proposed efficient algorithms based on successive convex approximation (SCA) and block coordinate descent (BCD) to optimize active beamforming at the transmitter, passive IRS beamforming, and the UAV's flight path. Studies such as \cite{7} and \cite{8} have focused on secure communications in UAV-assisted systems. In \cite{7}, a location-based beamforming strategy was proposed to improve system secrecy performance, leveraging the analytical form of secrecy outage probability. Meanwhile, \cite{8} addressed the maximization of secure energy efficiency in IRS-assisted UAV relaying by optimizing trajectory, IRS phase shifts, user scheduling, and transmit power, again employing SCA and BCD to handle the non-convexity of the problem.

However, a foundational assumption in nearly all existing work is that the transmitted signals follow an idealized Gaussian distribution \cite{4,5,6,7,8,9,PAMA}, which remarkably simplifies the analysis and optimization. This assumption, while mathematically convenient, is detached from practical communication systems, which transmit symbols from finite-alphabet constellations. Ignoring this mismatch can result in system designs that are theoretically sound but practically suboptimal, leading to inefficient beamforming, misleading trajectory paths, and degraded performance. This paper takes the crucial step of dropping the Gaussian input assumption and directly modeling the realistic finite-alphabet structure of transmitted signals.

Yet, releasing the Gaussian assumption significantly complicates the system design. Unlike the Gaussian case, where mutual information admits clean, convex-friendly expressions, the mutual information with finite-alphabet inputs is highly non-convex and lacks closed-form structure. Moreover, it introduces strong coupling between the precoder and UAV trajectory, making joint optimization extremely challenging. 

To address this challenge, this paper develops a new framework for UAV-assisted relay systems that explicitly accounts for finite-alphabet inputs in both trajectory and precoder design. By formulating a joint optimization problem over the UAV path and transmit strategies, and developing a tailored algorithm that combines alternating optimization (AO) with SCA, we provide a tractable solution that aligns communication system design more closely with practical signal structures.

The main contributions of this work are summarized below:
\begin{itemize}[leftmargin=0.4cm]
	\item Unlike most prior works that assume Gaussian inputs, which diverge from practical modulation schemes, this paper models the actual discrete nature of wireless signals. To the best of our knowledge, this is the first work to jointly optimize UAV trajectory and precoding under finite-alphabet input constraints in a multiple-input and multiple-output (MIMO) system, addressing a critical gap between theoretical design and real-world deployment.
	\item The finite-alphabet model results in a highly non-convex and coupled optimization problem due to the structure of the mutual information and variable dependencies. We propose an efficient algorithmic framework based on AO and SCA that enables tractable and effective joint optimization of the UAV's trajectory and the transmitter precoders.
\end{itemize} 

\section{Problem Formulation}
\begin{figure}[t]
  \centering
  \includegraphics[width=0.5\columnwidth]{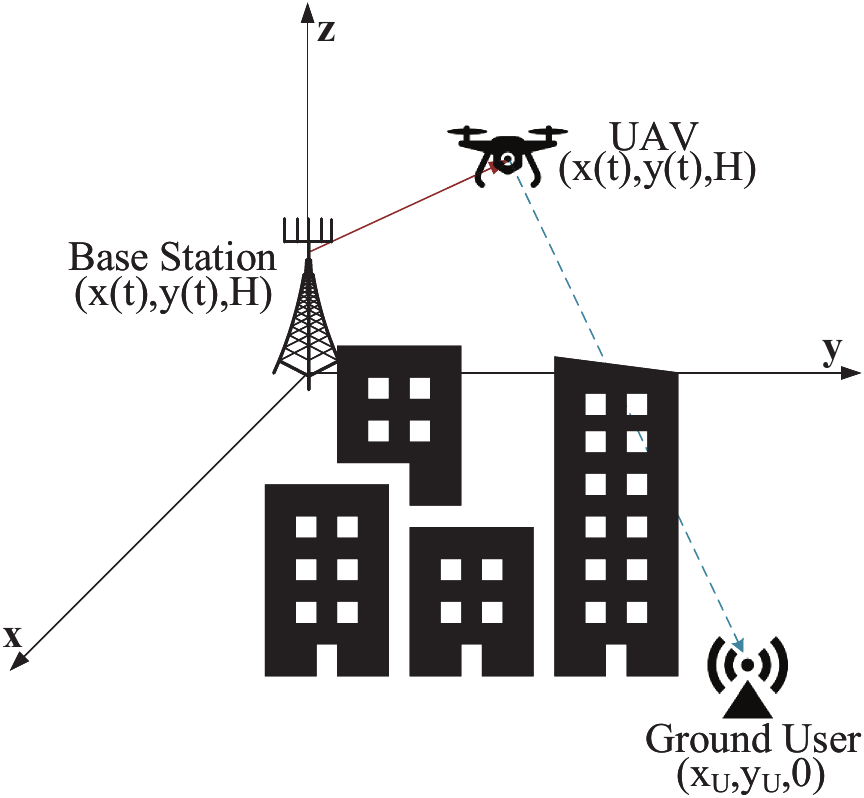}\\
  \caption{The MIMO UAV-assisted DF relay system comprising a UAV, a ground BS, and a GU.}
\label{Fig.1}
\end{figure}

We consider a MIMO UAV-assisted decode-and-forward (DF) relay system comprising a UAV, a ground base station (BS), and a ground user (GU), equipped with $\emph{N}_t$, $\emph{N}_u$, and $\emph{N}_r$ antennas, respectively, as depicted in Fig. \ref{Fig.1}.
Due to the presence of obstacles between the BS and the GU, a direct communication link is unavailable, necessitating the UAV's role as a relay node to facilitate information transfer between the BS and the GU. 

\subsection{System Model}
We adopt a 3D Cartesian coordinate system and assume that the UAV maintains a constant altitude of $\emph{H}$ throughout a finite time span $\emph{T}$.  For clarity, let $\mathbf{w}(\emph{t})=[\emph{x}(\emph{t}),\emph{y}(\emph{t})]^{\top},0\leq\emph{t}\leq\emph{T}$, denote the position of the UAV at time $\emph{t}$, where the initial position is $\mathbf{w}(0)=\mathbf{w}_{I}=[\emph{x}_{I},\emph{y}_{I}]^{\top}$ and the final position is $\mathbf{w}(\emph{T})=\mathbf{w}_{F}=[\emph{x}_{F},\emph{y}_{F}]^{\top}$. The BS and the GU are located at $\mathbf{p}_{B}=[\emph{x}_{B},\emph{y}_{B}]^{\top}$ and $\mathbf{p}_{U}=[\emph{x}_{U},\emph{y}_{U}]^{\top}$, respectively. Additionally, $\mathbf{v}(\emph{t})$ and $\mathbf{a}(\emph{t})$ are used to denote
the UAV's velocity and acceleration at time $\emph{t}$, respectively.

The time span $\emph{T}$ is divided into $\emph{N}+2$ sufficiently small time slots, each with a duration of ${\delta}_\emph{t}$. This partitioning allows for a reasonable assumption that within each time slot, the UAV's position, velocity, and acceleration remain constant.
The Euclidean distance between the BS and UAV and the Euclidean distance between the UAV and the GU are $\emph{d}_\emph{B}(\emph{n})=\sqrt{\Vert\mathbf{w}(\emph{n})-\mathbf{p}_{B}\Vert^{2}+\emph{H}^{2}}$ and $\emph{d}_\emph{U}(\emph{n})=\sqrt{\Vert\mathbf{w}(\emph{n})-\mathbf{p}_{U}\Vert^{2}+\emph{H}^{2}}$, respectively, where $\mathbf{w}(\emph{n})=\mathbf{w}(\emph{n}{\delta}_\emph{t}),\emph{n}=0,1,...,\emph{N}+1$.

In this paper, we employ the DF relay protocol, resulting in a two-phase communication process. In the first phase, the BS transmits data to the UAV, and in the second phase, the UAV forwards the received data to the GU. The BS-UAV channel, denoted by $\mathbf{H}_B\in\mathbb{C}^{\emph{N}_u \times \emph{N}_t}$, and the UAV-GU channel, denoted by $\mathbf{H}_U\in\mathbb{C}^{\emph{N}_r \times \emph{N}_u}$, are modeled as Rician block fading channels, which account for both line-of-sight (LoS) and non-line-of-sight (NLoS) components \cite{10}. Additionally, we assume effective compensation of the Doppler effect at both the UAV and GU. The two channels can be written as
\begin{equation} \label{eq 1}
	\begin{split}
		\mathbf{H}_\emph{B}(\emph{n})=&\sqrt{\rho_{0}\emph{d}^{-2}_\emph{B}(\emph{n})}[\sqrt{\frac{K}{K+1}}\mathbf{H}^{LoS}_\emph{B}(\emph{n})+\\&\sqrt{\frac{1}{K+1}}\mathbf{H}^{NLoS}_\emph{B}(\emph{n})], \ \emph{n}=0,1,...,\emph{N}+1,
	\end{split}
\end{equation}
\begin{equation} \label{eq 2}
	\begin{split}
		\mathbf{H}_\emph{U}(\emph{n})=&\sqrt{\rho_{0}\emph{d}^{-2}_\emph{U}(\emph{n})}[\sqrt{\frac{K}{K+1}}\mathbf{H}^{LoS}_\emph{U}(\emph{n})+\\&\sqrt{\frac{1}{K+1}}\mathbf{H}^{NLoS}_\emph{U}(\emph{n})], \ \emph{n}=0,1,...,\emph{N}+1,
	\end{split}
\end{equation}
where ${\rho}_{0}$ represents the channel power gain at a reference distance $\emph{d}_{0}=1$m, and $\emph{K}$ represents the Rician factor. $\mathbf{H}^{LoS}_\emph{B}(\emph{n})$ and $\mathbf{H}^{LoS}_\emph{U}(\emph{n})$ are the LoS components of the BS-UAV and UAV-GU links, respectively. $\mathbf{H}^{NLoS}_\emph{B}(\emph{n})$ and $\mathbf{H}^{NLoS}_\emph{U}(\emph{n})$ are the NLoS components of the BS-UAV and UAV-GU links, respectively, which follow the Rayleigh fading model.

In contrast to the commonly used Gaussian signals, this paper employs practical finite-alphabet signals as inputs. Specifically, let $\mathbf{x}_\emph{B}(\emph{n})\in\mathbb{C}^{\emph{N}_t \times 1}$ and $\mathbf{x}_\emph{U}(\emph{n})\in\mathbb{C}^{\emph{N}_u \times 1}$ denotes the signal transmitted at the BS and UAV at time slot $\emph{n}$, respectively, and each element of $\mathbf{x}_\emph{B}(\emph{n})$ and $\mathbf{x}_\emph{U}(\emph{n})$ is selected from an equiprobable discrete constellation with unit covariance matrix, i.e., $\mathrm{E}(\mathbf{x}_\emph{B}(\emph{n})\mathbf{x}_\emph{B}(\emph{n})^{H})=\mathbf{I}$ and $\mathrm{E}(\mathbf{x}_\emph{U}(\emph{n})\mathbf{x}_\emph{U}(\emph{n})^{H})=\mathbf{I}$, where $\mathbf{I}$ denotes the
identity matrix. The signals received at the UAV and the GU are given by 
\begin{equation} \label{eq 3}
	\mathbf{y}_\emph{U}(\emph{n})=\mathbf{H}_\emph{B}(\emph{n})\mathbf{P}_\emph{B}(\emph{n})\mathbf{x}_\emph{B}(\emph{n})+\mathbf{n}_\emph{U}(\emph{n}),
\end{equation}
\begin{equation} \label{eq 4}
	\mathbf{y}_\emph{G}(\emph{n})=\mathbf{H}_\emph{U}(\emph{n})\mathbf{P}_\emph{U}(\emph{n})\mathbf{x}_\emph{U}(\emph{n})+\mathbf{n}_\emph{G}(\emph{n}),
\end{equation}
where $\mathbf{P}_\emph{B}(\emph{n})\in\mathbb{C}^{\emph{N}_t \times \emph{N}_t}$ and $\mathbf{P}_\emph{U}(\emph{n})\in\mathbb{C}^{\emph{N}_u \times \emph{N}_u}$ are the precoding matrices at the BS and the UAV at time slot $\emph{n}$, respectively. $\mathbf{n}_\emph{U}(\emph{n})\in\mathbb{C}^{\emph{N}_u \times 1}$ and $\mathbf{n}_\emph{G}(\emph{n})\in\mathbb{C}^{\emph{N}_r \times 1}$ denote the additive white Gaussian noise (AWGN) in the BG-UAV link and the UAV-GU link at time slot $\emph{n}$, which satisfy $\mathbf{n}_\emph{U}(\emph{n})\sim\mathcal{CN} (0,\sigma_\emph{U}^{2}\mathbf{I})$ and $\mathbf{n}_\emph{G}(\emph{n})\sim\mathcal{CN} (0,\sigma_\emph{G}^{2}\mathbf{I})$, respectively. Under these premises, the information rate per Hertz between the BS and the UAV and between the UAV and the GU at time slot $\emph{n}$ can be expressed respectively as \cite{11}
\begin{equation} \label{eq 5}
	\begin{split}
		\bar{I}_\emph{U}  (\mathbf{x}_\emph{B}(\emph{n});&\mathbf{y}_\emph{U}(\emph{n})) =\emph{N}_t\mathrm{log}_{2}\rho-\\&\frac{1}{\rho^{\emph{N}_t}}\sum^{\rho^{\emph{N}_t}}_{m=1}
		\mathrm{E}_{\mathbf{n}_\emph{U}(\emph{n})} \left(\mathrm{log}_{2}\sum^{\rho^{\emph{N}_t}}_{k=1}\mathrm{exp}(-\emph{z}_\emph{Umk}(\emph{n}))\right),
	\end{split}
\end{equation}
\begin{equation} \label{eq 6}
	\begin{split}
		\bar{I}_\emph{G}  (\mathbf{x}_\emph{U}(\emph{n});&\mathbf{y}_\emph{G}(\emph{n})) =\emph{N}_u\mathrm{log}_{2}\rho-\\&\frac{1}{\rho^{\emph{N}_u}}\sum^{\rho^{\emph{N}_u}}_{m=1}
		\mathrm{E}_{\mathbf{n}_\emph{G}(\emph{n})} \left(\mathrm{log}_{2}\sum^{\rho^{\emph{N}_u}}_{k=1}\mathrm{exp}(-\emph{z}_\emph{Gmk}(\emph{n}))\right),
	\end{split}
\end{equation}
where $\rho$ is the size of the given equiprobable discrete constellations, and $\emph{z}_\emph{Umk}(\emph{n})=(\Vert \mathbf{H}_\emph{B}(\emph{n})\allowbreak\mathbf{P}_\emph{B}(\emph{n})\allowbreak\mathbf{u}_\emph{Bmk}(\emph{n})\allowbreak+\mathbf{n}_\emph{U}(\emph{n})\allowbreak \Vert^{2}\allowbreak-\Vert \mathbf{n}_\emph{U}(\emph{n}) \Vert^{2})/\sigma_\emph{U}^{2}$,   $\emph{z}_\emph{Gmk}(\emph{n})=(\Vert \mathbf{H}_\emph{U}(\emph{n})\mathbf{P}_\emph{U}(\emph{n})\mathbf{u}_\emph{Umk}(\emph{n})+\mathbf{n}_\emph{G}(\emph{n}) \Vert^{2}-\Vert \mathbf{n}_\emph{G}(\emph{n}) \Vert^{2})/\sigma_\emph{G}^{2}$. 
Here, $\mathbf{u}_\emph{Bmk}(\emph{n})=\mathbf{x}_\emph{Bm}(\emph{n})-\mathbf{x}_\emph{Bk}(\emph{n})$ and $\mathbf{u}_\emph{Umk}(\emph{n})=\mathbf{x}_\emph{Um}(\emph{n})-\mathbf{x}_\emph{Uk}(\emph{n})$, where
$\mathbf{x}_\emph{Bm}(\emph{n})$, $\mathbf{x}_\emph{Bk}(\emph{n})$, $\mathbf{x}_\emph{Um}(\emph{n})$, $\mathbf{x}_\emph{Uk}(\emph{n})$ are the input vectors taken independently from the $\rho$-ary discrete constellations.

Equations \eqref{eq 5} and \eqref{eq 6} can be further simplified as\cite{12}
\begin{equation} \label{eq 7}
	\begin{split}
		I_\emph{U}  (\mathbf{x}_\emph{B}&(\emph{n});\mathbf{y}_\emph{U}(\emph{n})) =2\emph{N}_t\mathrm{log}_{2}\rho-\\&\mathrm{log}_{2}\sum^{\rho^{\emph{N}_t}}_{m=1}\sum^{\rho^{\emph{N}_t}}_{k=1}\mathrm{exp}(-\frac{\Vert\mathbf{H}_\emph{B}(\emph{n})\mathbf{P}_\emph{B}(\emph{n})\mathbf{u}_\emph{Bmk}(\emph{n}) \Vert^{2}}{4\sigma_\emph{U}^{2}}),
	\end{split}
\end{equation}
\begin{equation} \label{eq 8}
	\begin{split}
		I_\emph{G}  (\mathbf{x}_\emph{U}&(\emph{n});\mathbf{y}_\emph{G}(\emph{n})) =2\emph{N}_u\mathrm{log}_{2}\rho-\\&\mathrm{log}_{2}\sum^{\rho^{\emph{N}_u}}_{m=1}\sum^{\rho^{\emph{N}_u}}_{k=1}\mathrm{exp}(-\frac{\Vert\mathbf{H}_\emph{U}(\emph{n})\mathbf{P}_\emph{U}(\emph{n})\mathbf{u}_\emph{Umk}(\emph{n}) \Vert^{2}}{4\sigma_\emph{G}^{2}}).
	\end{split}
\end{equation}

\subsection{Problem Formulation}
The average information rate per Hertz of the DF relay system can be written as
\begin{equation} \label{eq 9}
	\emph{R}_{avg} = \min(\emph{R}_{\emph{U}},\emph{R}_{\emph{G}}),
\end{equation}
where 
\begin{equation} \label{eq 10}
	\emph{R}_{\emph{U}} = \sum^{\emph{N+1}}_{n=0}\frac{{\delta}_\emph{t}I_\emph{U}  (\mathbf{x}_\emph{B}(\emph{n});\mathbf{y}_\emph{U}(\emph{n}))}{\emph{T}} ,	\emph{R}_{\emph{G}} = \sum^{\emph{N+1}}_{n=0}\frac{{\delta}_\emph{t}I_\emph{G}  (\mathbf{x}_\emph{U}(\emph{n});\mathbf{y}_\emph{G}(\emph{n}))}{\emph{T}}.
\end{equation}

At any time $\emph{t}$, we have
$\mathbf{v}(\emph{t})=\mathbf{w}^{\prime}(\emph{t})$ and $\mathbf{a}(\emph{t})=\mathbf{v}^{\prime}(\emph{t})$, where the superscript represents the first-order derivative.
By employing both first and second-order Taylor expansions, we have the following equations at time slot $\emph{n}$:
\begin{equation} \label{eq 11}
	\mathbf{w}(\emph{n}+1) = \mathbf{w}(\emph{n})+\mathbf{v}(\emph{n}){\delta}_\emph{t}+	\frac{1}{2}\mathbf{a}(\emph{n}){\delta}^{2}_\emph{t},
\end{equation}
\begin{equation} \label{eq 12}
	\mathbf{v}(\emph{n}+1) = \mathbf{v}(\emph{n})+\mathbf{a}(\emph{n}){\delta}_\emph{t}.
\end{equation}

The objective of this paper is to jointly optimize the precoders and the UAV trajectory to maximize the average information rate $\emph{R}_{avg}$, while adhering to transmit power constraints and UAV flight limitations. Accordingly, the formulated problem aims to achieve the optimal balance between efficient data transmission and practical system constraints, which is described as follows:
\begin{subequations} \label{eq 13}
	\begin{align}
		\operatorname*{max.}\limits_{\mathbf{w},\mathbf{v},\mathbf{a},\mathbf{P}_\emph{B},\mathbf{P}_\emph{U}} \ & \emph{R}_{avg} & \label{eq:subeq1}\\
		%	\{max.}_{\mathbf{w},\mathbf{v},\mathbf{a},\mathbf{m}}. & ~ & \xi\\
	%    \mathrm{s.~t.} & ~ &  \emph{R}_{\emph{U}} \geq \xi,\\\
	%    & ~ &  \emph{R}_{\emph{G}} \geq \xi,\\\
	\mathrm{s.~t.} \ &  \mathrm{Tr}(\mathbf{P}^{H}_\emph{B}(\emph{n})\mathbf{P}_\emph{B}(\emph{n}))\leq \emph{W}_{\emph{B}}, \ \forall \emph{n}, & \label{eq:subeq2}\\\
	\ &  \mathrm{Tr}(\mathbf{P}^{H}_\emph{U}(\emph{n})\mathbf{P}_\emph{U}(\emph{n}))\leq \emph{W}_{\emph{U}}, \ \forall \emph{n}, & \label{eq:subeq3}\\\
	\ &  	\mathbf{w}(0)=\mathbf{w}_\emph{I}, \  \mathbf{w}(\emph{N}+1)=\mathbf{w}_\emph{F}, & \label{eq:subeq4}\\\
	%    & ~ &   \mathbf{v}(0)=\mathbf{v}_\emph{I}, \  \mathbf{v}(\emph{N}+1)=\mathbf{v}_\emph{F},\\\
	\ &   \Vert \mathbf{v}(\emph{n}) \Vert \leq \emph{v}_{max}, \ \emph{n}=0,1,...,\emph{N}, & \label{eq:subeq5}\\\
	\ &   \Vert \mathbf{a}(\emph{n}) \Vert \leq \emph{a}_{max}, \ \emph{n}=1,2,...,\emph{N}, & \label{eq:subeq6}\\\
	%    & ~ &   \Vert \mathbf{v}(\emph{n}) \Vert \geq \emph{v}_{min}, \ \emph{n}=1,2,...,\emph{N},\\\
	\ &   \eqref{eq 11}, \ \eqref{eq 12}. \nonumber
%	& \label{eq:subeq7}
\end{align}
\end{subequations}
In problem \eqref{eq 13}, constraints \eqref{eq:subeq2} and \eqref{eq:subeq3} limit the transmit power at the BS and the UAV, with $\emph{W}_{\emph{B}}$
and $\emph{W}_{\emph{U}}$ representing the maximum transmit power at the BS and the UAV, respectively. Constraint \eqref{eq:subeq4} gives the initial and final positions of the UAV. The maximum velocity and acceleration of the UAV are indicated in constraints \eqref{eq:subeq5} and \eqref{eq:subeq6}, respectively. 

\section{Methodology}
This section focuses on solving \eqref{eq 13} and analyzing the influence of precoder design and UAV trajectory on the DF system rate. 
To address \eqref{eq 13}, we begin by introducing a slack variable $\xi$, which allows us to reformulate \eqref{eq 13} as follows:
\begin{subequations} \label{eq 14}
	\begin{align}
		\operatorname*{max.}\limits_{\mathbf{w},\mathbf{v},\mathbf{a},\mathbf{P}_\emph{B},\mathbf{P}_\emph{U},\xi} \ & \xi & \label{eq:subeq8} \\
		%	\{max.}_{\mathbf{w},\mathbf{v},\mathbf{a},\mathbf{m}}. & ~ & \xi\\
	\mathrm{s.~t.} \ & \sum^{\emph{N+1}}_{n=0}I_\emph{U}  (\mathbf{x}_\emph{B}(\emph{n});\mathbf{y}_\emph{U}(\emph{n})) \geq \xi, & \label{eq:subeq9} \\\
	\ & \sum^{\emph{N+1}}_{n=0}I_\emph{G}  (\mathbf{x}_\emph{U}(\emph{n});\mathbf{y}_\emph{G}(\emph{n})) \geq \xi, & \label{eq:subeq10} \\\
	%    & ~ &  \mathrm{Tr}(\mathbf{P}^{H}_\emph{B}(\emph{n})\mathbf{P}_\emph{B}(\emph{n}))\leq \emph{W}_{\emph{B}}, \ \emph{n}=0,1,...,\emph{N+1},\\\
	%    & ~ &  \mathrm{Tr}(\mathbf{P}^{H}_\emph{U}(\emph{n})\mathbf{P}_\emph{U}(\emph{n}))\leq \emph{W}_{\emph{U}}, \ \emph{n}=0,1,...,\emph{N+1},\\\
	%    & ~ &  	\mathbf{w}(0)=\mathbf{w}_\emph{I}, \  \mathbf{w}(\emph{N}+1)=\mathbf{w}_\emph{F},\\\
	%    & ~ &   \mathbf{v}(0)=\mathbf{v}_\emph{I}, \  \mathbf{v}(\emph{N}+1)=\mathbf{v}_\emph{F},\\\
	%    & ~ &   \Vert \mathbf{v}(\emph{n}) \Vert \leq \emph{v}_{max}, \ \emph{n}=1,2,...,\emph{N},\\\
	%    & ~ &   \Vert \mathbf{a}(\emph{n}) \Vert \leq \emph{a}_{max}, \ \emph{n}=0,1,...,\emph{N},\\\
	%    & ~ &   \Vert \mathbf{v}(\emph{n}) \Vert \geq \emph{v}_{min}, \ \emph{n}=1,2,...,\emph{N},\\\
	\ &  \eqref{eq:subeq2}-\eqref{eq:subeq6}, \ \eqref{eq 11}, \ \eqref{eq 12}. \nonumber
%	& \label{eq:subeq11}
\end{align}
\end{subequations}

It is important to note that the optimization variables are solely related to the second term of the information rate function (see \eqref{eq 7} and \eqref{eq 8}). On the other hand, the logarithmic nature of the function further complicates the resolution of problem \eqref{eq 14}. To address this complexity, we adopt a similar approach to that proposed in \cite{4}, transforming \eqref{eq 14} into a more manageable form as follows.
\begin{subequations} \label{eq 15}
	\begin{align}
		\ & \operatorname*{min.}\limits_{\mathbf{w},\mathbf{v},\mathbf{a},\mathbf{P}_\emph{B},\mathbf{P}_\emph{U},\xi} \xi & \label{eq:subeq12}\\
		%	\{max.}_{\mathbf{w},\mathbf{v},\mathbf{a},\mathbf{m}}. & ~ & \xi\\
	\ & \mathrm{s.~t.} \nonumber  \\  \ & \sum_{n=0}^{N+1}\sum^{\rho^{\emph{N}_t}}_{m=1}\sum^{\rho^{\emph{N}_t}}_{k=1}\mathrm{exp}(-\frac{\Vert\mathbf{H}_\emph{B}(\emph{n})\mathbf{P}_\emph{B}(\emph{n})\mathbf{u}_\emph{Bmk}(\emph{n}) \Vert^{2}}{4\sigma_\emph{U}^{2}}) \leq \xi+\emph{O}_\emph{B}, & \label{eq:subeq13}\\\
	\ &  \sum_{n=0}^{N+1}\sum^{\rho^{\emph{N}_u}}_{m=1}\sum^{\rho^{\emph{N}_u}}_{k=1}\mathrm{exp}(-\frac{\Vert\mathbf{H}_\emph{U}(\emph{n})\mathbf{P}_\emph{U}(\emph{n})\mathbf{u}_\emph{Umk}(\emph{n}) \Vert^{2}}{4\sigma_\emph{G}^{2}}) \leq \xi+\emph{O}_\emph{U}, & \label{eq:subeq14}\\\
	\ &  \eqref{eq:subeq2}-\eqref{eq:subeq6}, \ \eqref{eq 11}, \ \eqref{eq 12}, \nonumber
%	& \label{eq:subeq15}
\end{align}
\end{subequations}
where $\emph{O}_\emph{B}=2(N+2)\emph{N}_t\mathrm{log}_{2}\rho$ and $\emph{O}_\emph{U}=2(N+2)\emph{N}_u\mathrm{log}_{2}\rho$.

Handling problem \eqref{eq 15} is challenging due to the coupling of variable $\mathbf{w}$ with $\mathbf{P}_\emph{B}$ and $\mathbf{P}_\emph{U}$ in constraints \eqref{eq:subeq13} and \eqref{eq:subeq14}. To address this complexity, we develop an iterative optimization algorithm based on the AO framework to tackle \eqref{eq 15}. Specifically, the problem is decomposed into two sub-problems, allowing for more manageable optimization steps. The SCA technique is then applied to iteratively solve these sub-problems, providing a systematic approach to optimize the coupled variables and ultimately find a solution to the original problem.

\subsection{Optimization of the UAV trajectory}
To start with, we focus on optimizing the UAV's trajectory while maintaining fixed precoders. With this approach, the original problem can be reformulated as
\begin{subequations} \label{eq 16}
	\begin{align}
		\ & \operatorname*{min.}\limits_{\mathbf{w},\mathbf{v},\mathbf{a},\xi}  \xi & \label{eq:subeq16}\\
		%	\{max.}_{\mathbf{w},\mathbf{v},\mathbf{a},\mathbf{m}}. & ~ & \xi\\
	\mathrm{s.~t.}\ &   \sum_{n=0}^{N+1}\sum^{\rho^{\emph{N}_t}}_{m=1}\sum^{\rho^{\emph{N}_t}}_{k=1}\mathrm{exp}(-\frac{\rho_{0}\emph{A}_\emph{Bmk}(\emph{n})}{\Vert\mathbf{w}(\emph{n})-\mathbf{p}_{B}\Vert^{2}+\emph{H}^{2}})) \leq \xi+\emph{O}_\emph{B}, & \label{eq:subeq17}\\\
	\ &  \sum_{n=0}^{N+1}\sum^{\rho^{\emph{N}_u}}_{m=1}\sum^{\rho^{\emph{N}_u}}_{k=1}\mathrm{exp}(-\frac{\rho_{0}\emph{A}_\emph{Umk}(\emph{n})}{\Vert\mathbf{w}(\emph{n})-\mathbf{p}_{U}\Vert^{2}+\emph{H}^{2}})) \leq \xi+\emph{O}_\emph{U}, & \label{eq:subeq18}\\\
	\ &  \eqref{eq:subeq4}-\eqref{eq:subeq6}, \ \eqref{eq 11}, \ \eqref{eq 12}, \nonumber
%	& \label{eq:subeq19}
\end{align}
\end{subequations}
where $\emph{A}_\emph{Bmk}(\emph{n})=\frac{\Vert\mathbf{H}_\emph{rB}(\emph{n})\mathbf{P}_\emph{B}(\emph{n})\mathbf{u}_\emph{Bmk}(\emph{n}) \Vert^{2}}{4\sigma_\emph{U}^{2}}$, $\emph{A}_\emph{Umk}(\emph{n})=\frac{\Vert\mathbf{H}_\emph{rU}(\emph{n})\mathbf{P}_\emph{U}(\emph{n})\mathbf{u}_\emph{Umk}(\emph{n}) \Vert^{2}}{4\sigma_\emph{G}^{2}}$, $\forall \emph{n}$, and
\begin{equation*} \label{eq 19}
\mathbf{H}_\emph{rB}(\emph{n}) = \sqrt{\frac{K}{K+1}}\mathbf{H}^{LoS}_\emph{B}(\emph{n})+\sqrt{\frac{1}{K+1}}\mathbf{H}^{NLoS}_\emph{B}(\emph{n}), \ \forall \emph{n},
\end{equation*} 
\begin{equation*} \label{eq 20}
\mathbf{H}_\emph{rU}(\emph{n}) = \sqrt{\frac{K}{K+1}}\mathbf{H}^{LoS}_\emph{U}(\emph{n})+\sqrt{\frac{1}{K+1}}\mathbf{H}^{NLoS}_\emph{U}(\emph{n}), \ \forall \emph{n}.
\end{equation*} 

To handle the non-convexity of constraints \eqref{eq:subeq17} and \eqref{eq:subeq18}, we define a function
\begin{equation*}
\emph{F}_{sj}(\emph{n})\triangleq\sum^{\rho^{\emph{N}_s}}_{m=1}\sum^{\rho^{\emph{N}_s}}_{k=1}e^{-D_{jmk}(\emph{n})}, s\in\{t,u\}, j\in\{B,U\}, \ \forall \emph{n},
\end{equation*}
where $D_{jmk}(\emph{n})=\rho_{0}\emph{A}_\emph{jmk}(\emph{n})/(\Vert\mathbf{w}(\emph{n})-\mathbf{p}_{j}\Vert^{2}+\emph{H}^{2})$.

\begin{prop} \label{prop 1}
$F_{sj}({n})$ can be convexly approximated as
\begin{equation*}
	F^{lb}_{sj}({n})=\sum^{\rho^{\emph{N}_s}}_{m=1}\sum^{\rho^{\emph{N}_s}}_{k=1}e^{-D^{lb}_{jmk}({n})}, s\in\{t,u\}, j\in\{B,U\}, \ \forall {n},
\end{equation*}
\begin{equation*}
\begin{split}
	&D^{lb}_{jmk}({n})= \\& \frac{\rho_{0}\emph{A}_\emph{jmk}({n})}{\Vert\mathbf{w}^{(i)}({n})-\mathbf{p}_{j}\Vert^{2}+\emph{H}^{2}}-\frac{\rho_{0}\emph{A}_\emph{jmk}({n})}{(\Vert\mathbf{w}^{(i)}({n})-\mathbf{p}_{j}\Vert^{2}+\emph{H}^{2})^{2}}\\& \cdot(\Vert\mathbf{w}({n})-\mathbf{p}_{j}\Vert^{2}-\Vert\mathbf{w}^{(i)}({n})-\mathbf{p}_{j}\Vert^{2}), j\in\{B,U\}, \forall {n},
\end{split}
\end{equation*}
where $\mathbf{w}^{(i)}({n}), \forall {n}$, denotes the trajectory solution obtained in the $(i-1)$-th iteration of the SCA procedure.
\end{prop}

\begin{Proof}
Let $y_{j}(\emph{n})=\Vert\mathbf{w}(\emph{n})-\mathbf{p}_{j}\Vert^{2}$ and $y^{(i)}_{j}(\emph{n})=\Vert\mathbf{w}^{(i)}(\emph{n})-\mathbf{p}_{j}\Vert^{2}, \ j=B,U, \ \forall \emph{n}.$ Then, the first-order Taylor approximation of function $D_{jmk}(\emph{n})$ at a given point $y^{(i)}_{j}(\emph{n})$ is
\begin{equation*}
	\begin{split}
		D_{jmk}(\emph{n}) &\geq \frac{\rho_{0}\emph{A}_\emph{jmk}(\emph{n})}{y^{(i)}_{j}(\emph{n})+\emph{H}^{2}}-\frac{\rho_{0}\emph{A}_\emph{jmk}(\emph{n})}{(y^{(i)}_{j}(\emph{n})+\emph{H}^{2})^{2}}(y_{j}(\emph{n})-y^{(i)}_{j}(\emph{n}))\\&=D^{lb}_{jmk}(\emph{n}), \ j=B,U, \ \forall \emph{n}.
	\end{split}
\end{equation*}

$D^{lb}_{jmk}(\emph{n})$ is a concave function with respective to $\mathbf{w}(\emph{n})$. Thus, $\emph{F}^{lb}_{sj}(\emph{n})$ is a convex function with respective to $\mathbf{w}(\emph{n})$. Moreover, the first-order derivatives of $\emph{F}_{sj}(\emph{n})$ and $F^{lb}_{sj}(\emph{n})$ with respect to $\mathbf{w}(\emph{n})$ can be given by
\begin{equation} \label{eq 25}
	\begin{split}
		F^{\prime}_{sj}(\emph{n})&= \sum^{\rho^{\emph{N}_s}}_{m=1}\sum^{\rho^{\emph{N}_s}}_{k=1}(-\frac{2\rho_{0}\emph{A}_\emph{jmk}(\emph{n})\Vert\mathbf{w}(\emph{n})-\mathbf{p}_{j}\Vert}{(\Vert\mathbf{w}(\emph{n})-\mathbf{p}_{j}\Vert^{2}+\emph{H}^{2})^{2}})\cdot\\&\mathrm{exp}(-D_{jmk}(\emph{n})), s=t,u, \  j=B,U, \ \forall \emph{n},
	\end{split} 
\end{equation}
\begin{equation} \label{eq 26}
	\begin{split}
		F^{lb'}_{sj}(\emph{n})&=\sum^{\rho^{\emph{N}_s}}_{m=1}\sum^{\rho^{\emph{N}_s}}_{k=1}(-\frac{2\rho_{0}\emph{A}_\emph{jmk}(\emph{n})\Vert\mathbf{w}(\emph{n})-\mathbf{p}_{j}\Vert}{(\Vert\mathbf{w}^{(i)}(\emph{n})-\mathbf{p}_{j}\Vert^{2}+\emph{H}^{2})^{2}})\cdot \\& \mathrm{exp}(-D^{lb}_{jmk}(\emph{n}), s=t,u, \  j=B,U, \ \forall \emph{n}.
	\end{split} 
\end{equation}
Notably, \eqref{eq 25} and \eqref{eq 26} are equal when $\mathbf{w}(\emph{n})=\mathbf{w}^{i}(\emph{n}), \ \forall n$. Consequently,  function $F_{sj}(\emph{n})$ can be convexly approximated as $F^{lb}_{sj}(\emph{n})$. This proves Proposition \ref{prop 1}\cite{13}.
\end{Proof}

Based on proposition 1, by superseding $F_{tB}(\emph{n})$ and $F_{uU}(\emph{n})$ with $F^{lb}_{tB}(\emph{n})$ and $F^{lb}_{uU}(\emph{n}), \forall \emph{n}$, respectively, the non-convexity of \eqref{eq 16} with respect to $\mathbf{w}(\emph{n})$ is eliminated. Then, \eqref{eq 16} can be reformulated as the following form: 
\begin{subequations} \label{eq 27}
	\begin{align}
		\operatorname*{min.}\limits_{\mathbf{w},\mathbf{v},\mathbf{a},\xi} \ & \xi  & \label{eq:subeq20}\\
		%	\{max.}_{\mathbf{w},\mathbf{v},\mathbf{a},\mathbf{m}}. & ~ & \xi\\
	\mathrm{s.~t.} \ &  \sum_{n=0}^{N+1}\sum^{\rho^{\emph{N}_t}}_{m=1}\sum^{\rho^{\emph{N}_t}}_{k=1}\mathrm{exp}(-D^{lb}_{Bmk}(\emph{n})) \leq \xi+\emph{O}_\emph{B},& \label{eq:subeq21}\\\
	\ &  \sum_{n=0}^{N+1}\sum^{\rho^{\emph{N}_u}}_{m=1}\sum^{\rho^{\emph{N}_u}}_{k=1}\mathrm{exp}(-D^{lb}_{Umk}(\emph{n})) \leq \xi+\emph{O}_\emph{U},& \label{eq:subeq22}\\\
	\ &  \eqref{eq:subeq4}-\eqref{eq:subeq6}, \ \eqref{eq 11}, \ \eqref{eq 12}, \nonumber
%	& \label{eq:subeq23}
\end{align}
\end{subequations}
which is convex. By solving it iteratively using convex optimization toolboxes like CVX to yield the local point for the next iteration until convergence, the sub-optimal solution for problem \eqref{eq 16} is acquired. 

\subsection{Optimization of the precoders}
Next, we consider the optimization of precoders $\mathbf{P}_\emph{B}$ and $\mathbf{P}_\emph{U}$ with a given $\mathbf{w}$. The optimization problem can be rewritten as:
\begin{subequations} \label{eq 28}
	\begin{align}
		\operatorname*{min.}\limits_{\mathbf{P}_\emph{B},\mathbf{P}_\emph{U},\xi} \ & \xi & \label{eq:subeq24}\\
		%	\{max.}_{\mathbf{w},\mathbf{v},\mathbf{a},\mathbf{m}}. & ~ & \xi\\
	\mathrm{s.~t.} \ &  \sum_{n=0}^{N+1}\sum^{\rho^{\emph{N}_t}}_{m=1}\sum^{\rho^{\emph{N}_t}}_{k=1}\mathrm{exp}(-f_{BU}) \leq \xi+\emph{O}_\emph{B}, & \label{eq:subeq25}\\\
	\ &  \sum_{n=0}^{N+1}\sum^{\rho^{\emph{N}_u}}_{m=1}\sum^{\rho^{\emph{N}_u}}_{k=1}\mathrm{exp}(-f_{UG}) \leq \xi+\emph{O}_\emph{U}, & \label{eq:subeq26}\\\
	\ &  \eqref{eq:subeq2}-\eqref{eq:subeq3}, \nonumber
%	& \label{eq:subeq27}
\end{align}
\end{subequations}
where $f_{bg}=\frac{\mathrm{Tr}(\mathbf{u}_\emph{bmk}(\emph{n})\mathbf{u}^{H}_\emph{bmk}(\emph{n})\mathbf{P}^{H}_\emph{b}(\emph{n})\mathbf{H}^{H}_\emph{b}(\emph{n})\mathbf{H}_\emph{b}(\emph{n})\mathbf{P}_\emph{b}(\emph{n}))}{4\sigma_\emph{g}^{2}}, b\in\{B,U\}$, $g\in\{U,G\}$.

To eliminate the non-convexity of \eqref{eq:subeq25} and \eqref{eq:subeq26}, we define another function: 
\begin{equation*}
	E_{sbg}(\emph{n})=\sum^{\rho^{\emph{N}_s}}_{m=1}\sum^{\rho^{\emph{N}_s}}_{k=1}e^{-\frac{C_{bmk}}{4\sigma_\emph{g}^{2}}},
\end{equation*} 
for $s\in\{t,u\}$, $b\in\{B,U\}$, $g\in\{U,G\}$, and $\forall \emph{n}$, where 
\begin{equation*}
	\begin{split}
	 &C_{bmk}(\emph{n})=\\&\mathrm{Tr}(\mathbf{u}_\emph{bmk}(\emph{n})\mathbf{u}^{H}_\emph{bmk}(\emph{n})\mathbf{P}^{H}_\emph{b}(\emph{n})\mathbf{H}^{H}_\emph{b}(\emph{n})\mathbf{H}_\emph{b}(\emph{n})\mathbf{P}_\emph{b}(\emph{n})), b=B,U.
	\end{split}
\end{equation*}

\begin{prop} \label{prop 2}
A convex approximation of $E_{sbg}(\emph{n})$ is
\begin{equation}
	E^{ap}_{sbg}(\emph{n})=\sum^{\rho^{\emph{N}_s}}_{m=1}\sum^{\rho^{\emph{N}_s}}_{k=1}e^{-\frac{C^{ap}_{bmk}}{4\sigma_\emph{g}^{2}}},
\end{equation}
for $s\in\{t,u\}$, $b\in\{B,U\}$, $g\in\{U,G\}$, $\forall \emph{n}$, and
\begin{equation*}
\begin{split}
	C^{ap}_{bmk}(\emph{n})=&\mathrm{Tr}(\mathbf{u}_\emph{bmk}(\emph{n})\mathbf{u}^{H}_\emph{bmk}(\emph{n})\mathbf{P}^{(i)H}_\emph{b}(\emph{n})\mathbf{H}^{H}_\emph{b}(\emph{n})\mathbf{H}_\emph{b}(\emph{n})\mathbf{P}_\emph{b}(\emph{n}))+ \\  &\mathrm{Tr}(\mathbf{H}^{H}_\emph{b}(\emph{n})\mathbf{H}_\emph{b}(\emph{n})\mathbf{P}^{(i)}_\emph{b}(\emph{n})\mathbf{u}_\emph{bmk}(\emph{n})\mathbf{u}^{H}_\emph{bmk}(\emph{n})\mathbf{P}^{H}_\emph{b}(\emph{n}))- \\
	&\mathrm{Tr}(\mathbf{u}_\emph{bmk}(\emph{n})\mathbf{u}^{H}_\emph{bmk}(\emph{n})\mathbf{P}^{(i)H}_\emph{b}(\emph{n})\mathbf{H}^{H}_\emph{b}(\emph{n})\mathbf{H}_\emph{b}(\emph{n})\mathbf{P}^{(i)}_\emph{b}(\emph{n})), \\& b=B,U, \ \forall \emph{n},
\end{split}
\end{equation*}
where $\mathbf{P}^{(i)}_\emph{B}(\emph{n})$ and $\mathbf{P}^{(i)}_\emph{U}(\emph{n}), \ \forall \emph{n}$, represent the precoder solutions in the $(i-1)$-th iteration of the SCA procedure. 
\end{prop}

\begin{Proof}
Performing first-order Taylor expansion on $C_{bmk}$ yields\cite{14}
\begin{equation} \label{eq 32}
	\begin{split}
		&\mathrm{Tr}(\mathbf{u}_\emph{bmk}(\emph{n})\mathbf{u}^{H}_\emph{bmk}(\emph{n})\mathbf{P}^{(i)H}_\emph{b}(\emph{n})\mathbf{H}^{H}_\emph{b}(\emph{n})\mathbf{H}_\emph{b}(\emph{n})\mathbf{P}_\emph{b}(\emph{n}))+ \\  &\mathrm{Tr}(\mathbf{H}^{H}_\emph{b}(\emph{n})\mathbf{H}_\emph{b}(\emph{n})\mathbf{P}^{(i)}_\emph{b}(\emph{n})\mathbf{u}_\emph{bmk}(\emph{n})\mathbf{u}^{H}_\emph{bmk}(\emph{n})\mathbf{P}^{H}_\emph{b}(\emph{n}))- \\
		&\mathrm{Tr}(\mathbf{u}_\emph{bmk}(\emph{n})\mathbf{u}^{H}_\emph{bmk}(\emph{n})\mathbf{P}^{(i)H}_\emph{b}(\emph{n})\mathbf{H}^{H}_\emph{b}(\emph{n})\mathbf{H}_\emph{b}(\emph{n})\mathbf{P}^{(i)}_\emph{b}(\emph{n}))\\
		&=C^{ap}_{bmk}(\emph{n}), \ b=B,U, \ \forall \emph{n}.
	\end{split}
\end{equation}
It is obvious that $C^{ap}_{bmk}(\emph{n})$ is a convex surrogate function with respect to $\mathbf{P}_\emph{b}(\emph{n})$,$\ \forall n, \  b=B,U$. Thus, $E^{ap}_{sbg}(\emph{n})$ is also a convex surrogate function with respect to $\mathbf{P}_\emph{b}(\emph{n})$,$\  s=t,u, \ b=B,U, \ g=U,G, \ \forall \emph{n}$.

Furthermore, applying a first-order derivative on $E_{sbg}(\emph{n})$ with respect to $\mathbf{P}_\emph{b}(\emph{n})$ yields
\begin{equation} \label{eq 33}
	\begin{split}
		&E^{\prime}_{sbg}(\emph{n})= \\& \sum^{\rho^{\emph{N}_s}}_{m=1}\sum^{\rho^{\emph{N}_s}}_{k=1}(-\frac{2\mathbf{H}^{H}_\emph{b}(\emph{n})\mathbf{H}_\emph{b}(\emph{n})\mathbf{P}_\emph{b}(\emph{n})\mathbf{u}_\emph{bmk}(\emph{n})\mathbf{u}^{H}_\emph{bmk}(\emph{n})}{4\sigma_\emph{g}^{2}}))\cdot \\
		&(\mathrm{exp}(-\frac{C_{bmk}(\emph{n})}{4\sigma_\emph{g}^{2}}), s=t,u, \ b=B,U, \ g=U,G, \ \forall \emph{n}.
	\end{split}
\end{equation}
Besides, the first-order derivative of $E^{ap}_{sbg}(\emph{n})$ with respect to $\mathbf{P}_\emph{b}(\emph{n})$ is
\begin{equation} \label{eq 34}
	\begin{split}
		&E^{ap'}_{sbg}(\emph{n})= \\& \sum^{\rho^{\emph{N}_s}}_{m=1}\sum^{\rho^{\emph{N}_s}}_{k=1}(-\frac{2\mathbf{H}^{H}_\emph{b}(\emph{n})\mathbf{H}_\emph{b}(\emph{n})\mathbf{P}^{(i)}_\emph{b}(\emph{n})\mathbf{u}_\emph{bmk}(\emph{n})\mathbf{u}^{H}_\emph{bmk}(\emph{n})}{4\sigma_\emph{g}^{2}}))\cdot \\
		&(\mathrm{exp}(-\frac{C^{ap}_{bmk}(\emph{n})}{4\sigma_\emph{g}^{2}}),  s=t,u, \ b=B,U, \ g=U,G, \ \forall \emph{n}.
	\end{split}
\end{equation}
Obviously, when $\mathbf{P}_\emph{j}(\emph{n})=\mathbf{P}^{(i)}_\emph{j}(\emph{n})$, we have $E^{\prime}_{sbg}(\emph{n})=E^{ap'}_{sbg}(\emph{n})$, $s=t,u, \ b=B,U, \ g=U,G, \ \forall \emph{n}$. Thus, $E^{ap}_{sbg}(\emph{n})$ is a convex approximation of $E_{sbg}(\emph{n}), \ s=t,u, \ b=B,U, \ g=U,G, \ \forall \emph{n}$. This proves Proposition \ref{prop 2} \cite{13}.
\end{Proof}

Let $E^{ap}_{tBU}(\emph{n})$ and $E^{ap}_{uUG}(\emph{n})$ substitute for $E_{tBU}(\emph{n})$ and $E_{uUG}(\emph{n})$,$\ \forall \emph{n}$, respectively. Problem \eqref{eq 28} can be transformed into the following form:
\begin{subequations} \label{eq 35}
	\begin{align}
		\operatorname*{min.}\limits_{\mathbf{P}_\emph{B},\mathbf{P}_\emph{U},\xi} \ & \xi & \label{eq:subeq28}\\
		%	\{max.}_{\mathbf{w},\mathbf{v},\mathbf{a},\mathbf{m}}. & ~ & \xi\\
	\mathrm{s.~t.} \ &  \sum_{n=0}^{N+1}\sum^{\rho^{\emph{N}_t}}_{m=1}\sum^{\rho^{\emph{N}_t}}_{k=1}\mathrm{exp}(-\frac{C^{ap}_{Bmk}(\emph{n})}{4\sigma_\emph{U}^{2}}) \leq \xi+\emph{O}_\emph{B}, & \label{eq:subeq29}\\\
	\ &  \sum_{n=0}^{N+1}\sum^{\rho^{\emph{N}_u}}_{m=1}\sum^{\rho^{\emph{N}_u}}_{k=1}\mathrm{exp}(-\frac{C^{ap}_{Umk}(\emph{n})}{4\sigma_\emph{G}^{2}}) \leq \xi+\emph{O}_\emph{U}, & \label{eq:subeq30}\\\
	\ &  \eqref{eq:subeq2}-\eqref{eq:subeq3}, \nonumber
%	& \label{eq:subeq31}
\end{align}
\end{subequations}
which is a convex problem. By solving it iteratively using CVX to yield a local point for the next iteration until convergence, the sub-optimal solution for problem \eqref{eq 28} is acquired.

\subsection{The overall optimization algorithm}
Our Finite-Alphabet-Aware Optimization (FAAO) framework is outlined in Algorithm \ref{alg:alg1}. First, problem \eqref{eq 27} is addressed iteratively to derive a sub-optimal UAV trajectory $\mathbf{w}$. Next, sub-optimal precoders $\mathbf{P}_\emph{B}$ and $\mathbf{P}_\emph{U}$ at the BS and the UAV are obtained by iteratively solving problem \eqref{eq 35}. The use of the SCA technique guarantees convergence for each of these steps. Moreover, thanks to the convergent properties of the AO technique, the entire optimization algorithm converges to a locally optimal solution for problem \eqref{eq 15}.

\begin{algorithm}[t] %%\hspace{0.5cm} \Statex \hspace{0.5cm}
	\caption{Finite-Alphabet-Aware Optimization Framework.}\label{alg:alg1}
	\begin{algorithmic}[1]
		\State Initialize $\mathbf{w}^0(\emph{n}),\mathbf{P}^0_\emph{B}(\emph{n}),\mathbf{P}^0_\emph{U}(\emph{n}), \emph{n}=0,1,...,\emph{N}+1$, and the tolerance $tol$. Set the iteration index $i=0$. Compute $\emph{R}^0_{avg}$ according to \eqref{eq 9} and \eqref{eq 10}. 
		\State {\bf{repeat}}
		\State With the given point [$\mathbf{P}^{i}_\emph{B}(\emph{n}),\mathbf{P}^{i}_\emph{U}(\emph{n})$], solve problem \eqref{eq 27} by CVX iteratively to acquire the solution $\mathbf{w}^{i+1}(\emph{n})$, $\emph{n}=0,1,...,\emph{N}+1$.
		\State With the newly obtained local point $\mathbf{w}^{i+1}(\emph{n})$, solve problem \eqref{eq 35} by CVX iteratively to acquire the solution 
		$[\mathbf{P}^{i+1}_\emph{B}(\emph{n}),\mathbf{P}^{i+1}_\emph{U}(\emph{n})]$, $\emph{n}=0,1,...,\emph{N}+1$.
		\State Compute $\emph{R}^{i+1}_{avg}$ according to \eqref{eq 9} and \eqref{eq 10}.
		\State Let  $\tau=\left| \emph{R}^{i+1}_{avg}-\emph{R}^{i}_{avg} \right|$.
		\State Update $i=i+1$.
		\State {\bf{until}} $\tau<tol$.
		\State Output solution [$\mathbf{w}^{\star}(\emph{n}),\mathbf{P}^{\star}_\emph{B}(\emph{n}),\mathbf{P}^{\star}_\emph{U}(\emph{n})$], $\forall\emph{n}$.
	\end{algorithmic}
	\label{alg1}
\end{algorithm}

\section{Simulation Results}
This section presents simulation results to demonstrate the efficacy of the proposed solution.
To ensure a fair comparison, we adopt the same parameter settings as that in \cite{4} and \cite{13}. Specifically, the numbers of antennas at the BS, the UAV, and the GU are set to $\emph{N}_t=\emph{N}_u=\emph{N}_r=2$. We discretize the communication period $\emph{T}$ with a time step size of  $\delta_\emph{t}=0.2$s. The noise power is $\sigma^{2}_\emph{B}=\sigma^{2}_\emph{U}=-120$ dBm, while the transmit power at the BS and the UAV are set to $\emph{W}_\emph{B}=\emph{W}_\emph{U}=\emph{W}$. 
Binary phase shift keying modulation is considered. 
The reference channel power is $\rho_{0}=-50$ dB, and the Rician factor $\emph{K}$ is set to $3$. The height of the UAV is set to be $100$m, and the initial and final positions of the UAV are $\mathbf{w}_{I}=[0,350]$ and $\mathbf{w}_{F}=[350,0]$, respectively.
Furthermore, the positions of the BS and GU are $\mathbf{p}_{B}=[0,0]^{\top},\mathbf{p}_{G}=[300,300]^{\top}$, respectively. $\emph{v}_{max}$ and $\emph{a}_{max}$ are set to be 100 m/s and 5 m/$\rm{s}^{2}$, respectively. 

\begin{figure}[t]
  \centering
  \includegraphics[width=0.7\columnwidth]{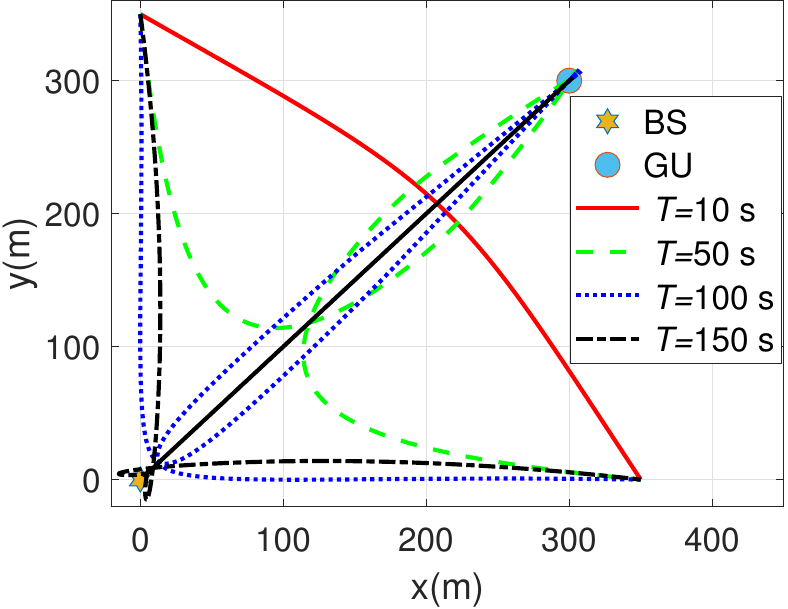}\\
  \caption{UAV trajectories of different communication period $\emph{T}$, $\emph{W}=20$ dBm.}
\label{Fig. 3}
\end{figure}

Fig. \ref{Fig. 3} depicts the UAV's flight trajectory for different values of $\emph{T}$ when $\emph{W}=20$ dBm. When time resources are extremely limited, such as $\emph{T}=10$ seconds, the UAV's trajectory is nearly a straight line from its initial position $\mathbf{w}_{I}$ to its final position $\mathbf{w}_{F}$, indicating a direct and time-efficient path. Conversely, when more time resources are available, such as $\emph{T}=100$ seconds or $\emph{T}=150$ seconds, the UAV demonstrates a tendency to approach either the BS or GU after departing from $\mathbf{w}_{I}$. This behavior enables the UAV to establish better channel conditions before eventually proceeding to $\mathbf{w}_{F}$, leveraging the extra time to optimize communication performance.

\begin{figure}[t]
  \centering
  \includegraphics[width=0.75\columnwidth]{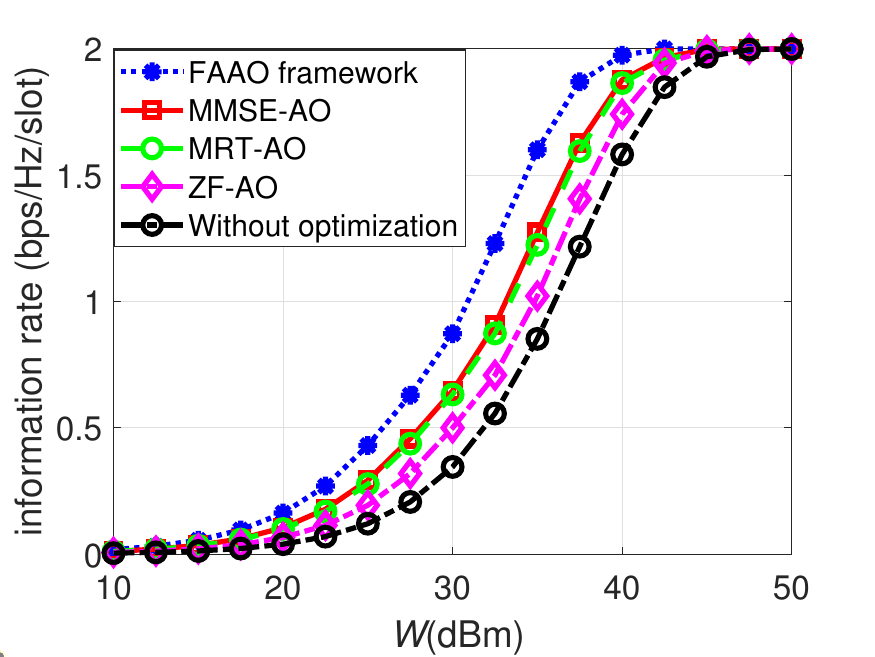}\\
  \caption{Information rate versus $\emph{W}$ under different schemes, $\emph{T}=50$ seconds.}
\label{Fig. 4}
\end{figure}

Fig. \ref{Fig. 4} illustrates the performance comparison among various schemes when $\emph{T}=50$ seconds. Three widely used precoding strategies under Gaussian inputs are compared: minimum mean square error (MMSE), maximum ratio transmission (MRT), and zero forcing (ZF). To the best of our knowledge,  there is no existing UAV trajectory optimization algorithm based on the 
MIMO achievable rate has been reported to date. Therefore, we employ our method to optimize UAV trajectory among all schemes. As shown, our FAAO framework outperforms the other schemes, demonstrating its effectiveness in practical systems, while also highlighting the significance of optimal designing optimization algorithms under finite-alphabet inputs.

\section{Conclusion}
This paper investigated the joint design of UAV trajectory and precoders in a relay system with finite-alphabet inputs. By moving beyond the conventional Gaussian input assumption, we addressed a more practical setting that reflects the use of discrete modulation schemes in real-world systems, enabling more accurate performance optimization and revealing the limitations of Gaussian-based designs.
% This paper investigated the joint design of UAV trajectory and precoders in a relay system with finite-alphabet inputs. By moving beyond the conventional Gaussian input assumption, we addressed a more practical and realistic setting that reflects the use of discrete modulation schemes in real-world systems. This shift enables more accurate performance optimization and reveals the limitations of existing Gaussian-based designs.
However, the removal of the Gaussian assumption introduces substantial challenges due to the non-convex and coupled nature of the resulting problem. To tackle this, we proposed an efficient algorithm based on AO and SCA, enabling tractable subproblem decomposition. Simulation results validate the effectiveness of our approach in practical systems, highlighting the significance of finite-alphabet-aware system design and offers a foundation for further research in realistic UAV-enabled communications.
% , showing consistent performance gains under practical signaling constraints. This work highlights the significance of finite-alphabet-aware system design and offers a foundation for further research in realistic UAV-enabled communications.

\bibliographystyle{IEEEtran}
\bibliography{mybibfile.bib}

\end{document}